# Step edge-mediated assembly of periodic arrays of long graphene nanoribbons on Au(111)

Chuanxu Ma,[§,†] Zhongcan Xiao,[‡] Wenchang Lu,[‡,⊥] Jingsong Huang,[§,⊥] Kunlun Hong,[§] J. Bernholc,[‡,⊥] An-Ping Li[§,*]

[§]Center for Nanophase Materials Sciences, Oak Ridge National Laboratory, Oak Ridge, TN 37831, USA

[†]The University of Tennessee, Knoxville, Tennessee 37996, USA

[‡]Department of Physics, North Carolina State University, Raleigh, NC 27695, USA

[⊥]Computational Sciences and Engineering Division, Oak Ridge National Laboratory, Oak Ridge, TN 37831, USA

*Email: apli@ornl.gov

**The influence of substrate steps on the bottom-up synthesis of atomically precise graphene nanoribbons (GNRs) on an Au(111) surface is investigated. A straight surface step is found to promote the assembly of long and compact arrays of polymers with enhanced interchain π−π stacking interaction, which creates a steric hindrance effect on cyclodehydrogenation to suppress the H passivation of polymer ends. The modified two-stage growth process results in periodic arrays of GNRs with doubled average length near step edges.**

On-surface synthesis of atomically precise graphene nanoribbons (GNRs) from rationally designed molecular precursors has attracted great attention.[1-7] In addition to the precise control of GNR widths and thus their electronic properties, long GNRs and periodic GNR arrays are the two long sought-after targets of GNR synthesis. First, long lengths are needed for fabrication of GNR devices and nanocircuits, although short-channel (20-nm) GNR field-effect transistors have been explored.[8] Second, well-aligned GNR arrays are desirable not only for device fabrication but also for optical and optoelectronic functionalities, such as polarized photoluminescence.[9-11] At elevated temperature, well-aligned GNR arrays can be further fused together into 2D graphene with ordered porous structures through lateral conjugations.[12] Significant progress has been made in addressing these goals. It was proposed that hydrogen passivation of polymer ends is a limiting factor for growing long GNRs because it can prevent further polymer growth.[13] Indeed, an effort to reduce the passivation effect by enlarging temperature separation between the polymerization and cyclodehydrogenation processes has led to longer ribbons.[14] Recently, GNR superlattice arrays with long ribbon lengths have been observed on Au(111) surface, which was attributed to the uniaxial anisotropy of the zigzag-patterned (22×√3) herringbone reconstruction.[15] Aligned GNR arrays were also achieved by using the heavily-stepped vicinal Au(788) surface[16,17] or reconstructed channels on the Au(110) as a template.[18] However, mechanistic understandings are still lacking on how the Au steps direct the formation of aligned GNR arrays and enhance the growth of long GNRs, and it remains unclear whether the Au step can help grow longer ribbons or form periodic arrays on a normal Au(111) surface. Here we address these questions and show that an enhanced π−π stacking at the Au(111) monoatomic step directs the assembly of polymers, and a steric hindrance at the step suppresses hydrogen passivation leading to periodic arrays of GNRs with significantly increased lengths.

We used 10,10'-dibromo-9,9'-bianthracene (DBBA) molecules to grow the seven-carbon-wide armchair GNRs (7-aGNRs), as illustrated in Fig. 1a (see ESI for experimental details), by following previous reports.[1, 19-21] Figure 1b shows a large-area STM image of 7-aGNRs after 470 and 670 K annealing. Four Au terraces with three nearly parallel monoatomic steps (marked as Steps 1 to 3) can be seen. Interestingly, near the straight step edges (marked with dashed white lines) the 7-aGNRs form quasi-periodic arrays, while those around the curved edges (marked with dashed black lines) are more randomly oriented. Periodic GNR arrays are particularly dominant on the two narrow terraces confined by three straight steps, which have widths of about 15 and 20 nm, respectively. These results indicate that a step edge can indeed direct the assembly of GNR periodic arrays. We find that the spatial separation between individual GNRs in the array varies due to repulsive interaction between the ribbon edges. As resolved in the atomic STM image of Fig. 1c, an average periodicity of about 1.7 nm is obtained, which can change between 1.5 and 2.1 nm. By controlling the coverage of the DBBA precursors during deposition, 7-aGNR arrays with different periodicities can be achieved, similarly to the previous report.[15] Not only can GNRs grow in parallel to the Au straight steps, they can grow directly on the step edges as well. As shown in Fig. 1d, a tilted GNR sits directly on the step edge, with the same atomic structure and electronic structure as those GNRs seen on the flat terrace.

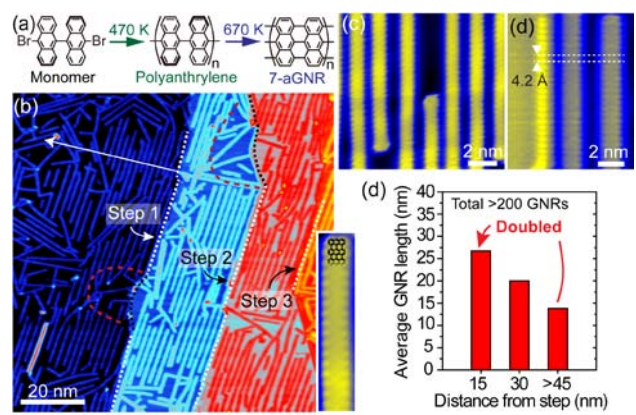

**Fig. 1** (a) Scheme of the bottom-up synthesis of 7-aGNRs from DBBA monomers via two-step annealing at 470 and 670 K, respectively. (b) Large-area STM image of GNR arrays in parallel to Au straight steps (setpoint: −2 V, 30 pA). White and black dashed lines mark the straight and curved edges in Steps 1-3, respectively. Red dashed curves enclose randomly oriented GNRs near curved step edges. Inset: High-resolution STM image of a single 7-aGNR superimposed with the structural model. (c) High-resolution STM image of the 7-aGNR periodic array (−0.2 V, 100 pA). (d) High-resolution STM image showing a 7-aGNR adsorbed on the upper edge of a step and two GNRs on the lower terrace (−1.0 V, 70 pA), with the same period along the ribbon. (e) Length distributions of the 7-aGNRs as a function of the lateral distance to straight Au steps. For GNRs unparallel to the steps, the position of the GNR center is used to measure the distance.

Besides the formation of periodic arrays, the GNRs near a straight step are generally longer when they are closer to the step, as shown along the white arrow line in Fig. 1b (see Fig. S1 for more examples). More precisely, we plot in Fig. 1e a statistical distribution of GNR lengths against the distance to nearby straight Au steps for over 200 GNRs, where the average length clearly increases as the distance to steps reduces. For example, the GNRs are

~27 nm long in regions within 15 nm from a step but are only ~13 nm long when being 45 nm away from a step. Therefore, the GNR lengths are almost doubled due to the proximity to the step.

In order to understand how the Au(111) steps facilitate the growth of periodic arrays of long GNRs, we examine individually the polymerization process and an intermediate process[22, 23] in the growth of the 7-aGNRs, as illustrated in Fig. 2a. After the polymerization, the polymer arrays are preferably aligned along the monoatomic steps, as shown in Fig. 2b. Furthermore, the polymer chains right next to the step are much longer in lengths as compared to those away from the steps. A high-resolution STM image of a polymer array near a step edge is shown as the inset of Fig. 2b. The separations between the polymer chains vary in the same array, which is smaller (8.8 Å) for chains closer to the Au step edge than for ordinary chains (10.8 Å) farther away from the step. The compact assembly of the polymers enhances the π−π stacking interaction between the polymers, which would significantly affect the reaction kinetics in the subsequent cyclodehydrogenation process.

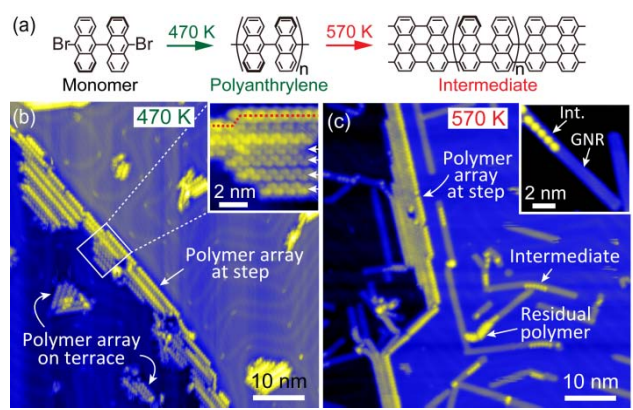

**Fig. 2** (a) Scheme of the temperature-dependent synthesis of the 7-aGNRs from DBBA monomers via an intermediate state. (b) Large-area STM image showing polymer arrays after 470 K annealing (−2 V, 10 pA). Note the Au step is {100}-type, according to the direction of the herringbone structures. Inset: High-resolution image of the box area in (b) (−2 V, 10 pA), where the white arrows mark four polymer chains. (c) Large-area STM image showing partially converted 7-aGNRs on the Au terraces and polymer arrays at the steps after 570 K annealing (−2 V, 40 pA). Inset: High-resolution image of the partially converted 7-aGNR with an intermediate (Int.) segment (−2 V, 40 pA).

The sample is then annealed at 570 K, lower than the normal graphitization temperature of 670 K. This treatment is known to generate an intermediate structure, which is a partially converted GNR with one side of the polyanthrylene converted to the GNR structure while the other side remains in the polymeric structure.[22-24] However, as shown in Fig. 2c, partially converted GNRs only occur at regions away from the steps, while the polymer arrays largely remain at the steps on the lower terrace side with a compact assembly. Therefore, the step-induced compact assembly is seen to suppress the cyclodehydrogenation, i.e., the temperature threshold of the polymer-to-GNR reaction is higher for a polymer at the step. This result seems to contradict the common belief that the Au step edges are usually active catalytic sites.[25, 26]

The compact polymer assembly is a key to understanding the observed suppression of cyclodehydrogenation at the step edge. We now focus on the high-resolution image of the polymer arrays shown in Fig. 3a. First, the atomic structures of two individual polymers comparably adsorbed on the step and the terrace are shown in Fig. 3b. Both display the same characteristic period of about 8.6 Å along the polymer, which is double of the 7-aGNR unit. On the other hand, the polymer chain adsorbed on the step edge is asymmetric across the width, while the one on the terrace is symmetric. The special adsorption configuration of the polymer on the step is well captured in the simulated STM image based on a structural model (Fig. 3c). The results indicate that a height difference for the two out-of-plane benzyne groups ($C_6H_4$) is responsible for the asymmetric features of the polymer on the step edge, which with graphitization will be converted to a tilted GNR on the step edge as seen in Fig. 1d. Second, an array of four polymer chains is shown in Fig. 3d as assembled on the lower terrace beside a step, and compared with another array assembled away from any steps (Fig. 3e). A separation distance is measured to be ca. 9.1 Å for polymers assembled near the step versus 11.2 Å for those assembled on the terrace. In the array beside the step, an even smaller (~8.1 Å) separation is seen between the polymers immediately adjacent to the step. These observed features can be well reproduced in the simulation results by using configuration models shown in Fig. 3f and 3g, respectively, for polymers assembled at the step and on the terrace.

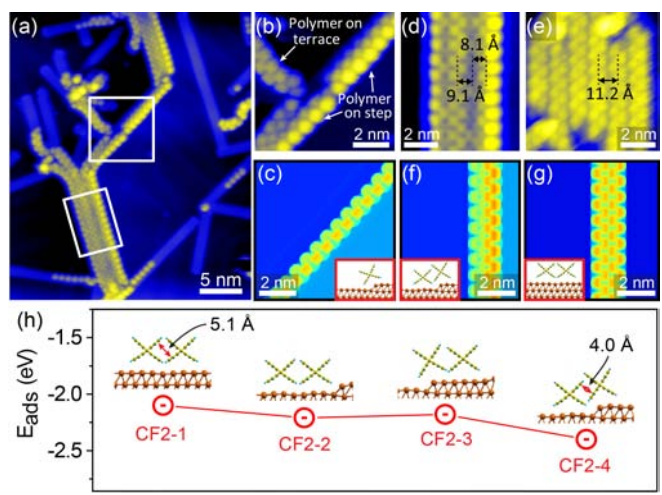

**Fig. 3** (a) STM image acquired in the same region as Fig. 2c with polymer arrays at the steps (−2 V, 40 pA). (b) Close-up STM image from the square box region in (a) (−2 V, 40 pA). (c) Simulated STM image of a single polymer adsorbed at the Au step. (d) Close-up STM image from the rectangle box region in (a) (−2 V, 40 pA). (e) Close-up STM image of a polymer array on the Au terrace formed after 470 K annealing (−2 V, 100 pA). (f) Simulated STM image of two polymer chains assembled at the Au step and on the lower terrace, respectively. (g) Simulated STM image of two polymers assembled on the Au terrace. Insets of (c), (f) and (g): Corresponding side-view structural models used for simulating the STM images. (h) Calculated adsorption energies $E_{ads}$ of polymer pairs assembled on the Au terrace or at the step, with the corresponding side-view structural models as insets. The π–π stacking distances in CF2-1 and CF2-4 are labeled and marked with double-arrowed lines.

We now calculate the adsorption energies $E_{ads}$ for pairs of polymers in different adsorption configurations as shown in Fig. 3h. The adsorption energies for polymer pairs assembled on the Au terrace (CF2-1), beside the step (CF2-2), and with one chain directly on the step and the other on the upper (CF2-3) or lower terrace (CF2-4), are −2.10, −2.21, −2.18, and− 2.40 eV, respectively. In comparison, the calculated adsorption energies $E_{ads}$ are around −1.00 eV for single polymers on the surface, and a π−π interaction energy of −0.37 eV is found for two neighboring polymer chains in vacuum. Clearly, these polymers prefer to assembly into arrays rather than being "isolated" single polymers, and the π−π stacking interaction in the polymer assemblies enhances the adsorption energies. The enhancement of π−π stacking interaction for the CF2-4 is the strongest as the neighboring benzyne groups are closest (4.0 Å) across the two polymers. Therefore, the CF2-4 is the most energetically preferable configuration when polymers assemble on the Au surface, giving rise to polymer arrays along the lower terrace. Interestingly, the different strengths of π−π interactions are reflected in the change of dihedral angles between neighboring anthracene units in the polymer pairs (see Fig. S2 and discussions in the ESI).

On the basis of all presented results, we are ready to comment on the effect of Au steps on GNR growth. During the polymerization process, polymer segments can easily move to assemble along the steps and form arrays. This compact configuration can spread several chains away from the step, and then transform to the ordinary assembly configuration, as observed in Fig. 2b. Previous reports have demonstrated that the initial step of the neighboring anthrylene units rotating and forming a single C-C bond is the rate-limiting step in the cyclodehydrogenation.[22, 27] The compact assembly of polymers enhances steric hindrance, which will significantly increase the energy barrier and result in a higher temperature threshold for the polymer-to-GNR reactions. For the single polymer chains adsorbed right on the step edge, the cyclodehydrogenation is also suppressed (Fig. 3b), which can be attributed to a different mechanism of the enhanced polymer-substrate interaction that stabilizes the cyclodehydrogenation starting point. The dispersive forces effect is known to be critical with lower coordination, such as the Au step.[28]

It was previously suggested that a larger temperature separation between the polymerization and the cyclodehydrogenation can result in longer polymers, and thus longer GNRs.[14] If the temperature difference is small, the cyclodehydrogenation might already happen during the polymerization annealing, and the generated excess hydrogen can passivate the radical ends of the polymers and thereby halt the growth of longer polymers.[13] Since the applied polymerization temperature is much higher than the recombinative desorption temperature (111 K) of adsorbed H atoms from Au(111) surface,[29] the adsorbed H atoms will either recombine as $H_2$ and desorb from the substrate surface or passivate the very nearby polymers on the terrace. We have showed that the polymers at the steps are indeed longer overall than those on the terraces, which gives rise to long GNR arrays after further annealing. Note, we find that a relatively low deposition rate is needed to harness the step edge-mediated effect for growing periodic and long GNR arrays (Fig. S3). Moreover, if the coverage of GNRs is too low (less than 0.4 monolayer), the weak repulsive interactions between the GNRs will either lead to a random arrangement (as seen in Fig. 2c) or be overwhelmed by the template effect of the uniaxially anisotropic zigzag pattern of the herringbone structures,[15] which can also give rise to a uniform separation (Fig. S4).

In summary, we investigate the role of the Au steps in the bottom-up synthesis of atomically precise GNRs. The monoatomic steps on Au(111) are found to direct the assembly of polymers and to increase the threshold of the cyclodehydrogenation temperature due to an enhanced π−π stacking, resulting in GNR periodic arrays with long ribbon lengths. Our findings uncover a route to make periodic and long GNR arrays on a normal Au(111) surface. As monoatomic steps are ubiquitous on metal surfaces, the mechanisms revealed here should also offer insights into GNR synthesis on other metal surfaces.

A portion of this research was conducted at the Center for Nanophase Materials Sciences (CNMS), which is a DOE Office of Science User Facility. This research was funded by grants ONR N00014-16-1-3213 and N00014-16-1-3153. The development of the RMG code was funded by NSF grant OAC-1740309. Supercomputer time was provided by NSF grant ACI-1615114 at the National Center for Supercomputing Applications (NSF OCI-0725070 and ACI-1238993).

**Conflicts of interest**

There are no conflicts to declare.

**Notes and references:**

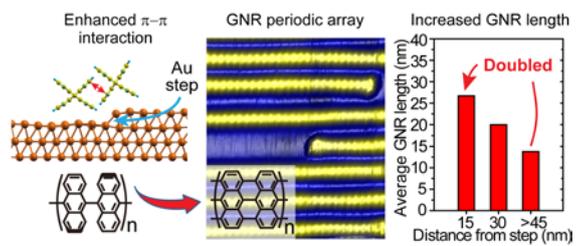

For TOC only.

Supporting Information

# Step edge-mediated assembly of periodic arrays of long graphene nanoribbons on Au(111)

Chuanxu Ma, Zhongcan Xiao, Wenchang Lu, Jingsong Huang, Kunlun Hong, Jerzy Bernholc, An-Ping Li

**Experimental details:**

The Au(111) single crystal was cleaned by repeated cycles of $Ar^+$ sputtering and annealing at 740 K. DBBA molecules with a purity of 98.7% were degassed and then evaporated onto the Au substrate at a temperature of 470 K with one monolayer coverage. At a constant source-substrate distance and background pressure (about $2\times10^{-10}$ torr), the source cell temperature was varied with deposition pressures of $6\times10^{-9}$ and $2\times10^{-8}$ torr to achieve low and high deposition rates, respectively. The sample was subsequently annealed at 470 K for 10 min and 670 K for 20 min, respectively, for polymerization and cyclodehydrogenation to form the 7-aGNRs. To grow the intermediate state, a lower graphitization temperature of 570 K was employed, similarly to previous reports.[22-24] The as-grown GNR sample was subsequently transferred from the growth chamber to the STM chamber under ultrahigh vacuum (UHV). The STM characterizations were performed with a home-made variable temperature system at 105 K with a clean commercial PtIr tip. All STM images were acquired in a constant-current mode. The bias voltage was applied to the sample bias with respect to the tip.

**DFT calculation details:**

The atomic structures and electronic properties of polymer assemblies on Au substrate were calculated with density functional theory (DFT). The calculations were performed with Quantum Espresso package,[30] using ultrasoft pseudopotentials[31] and Perdew-Burke-Ernzerhof exchange-correlation functional,[32] which has a good performance in modeling metal substrate.[33] A non-local vdw-df functional[34] was used to achieve higher accuracy in the simulation of van der Waals interaction between polymer and Au,[35] which is shown to be a good choice to produce accurate adsorption energies for graphene on Au,[36] especially on substrate steps.[28] The energy cutoffs were 30 Ry and 300 Ry for the wavefunctions and charge density, respectively. The atomic structures were relaxed until forces reached a threshold of 0.002 RyÅ$^{-1}$. The iso-current STM images were simulated based on Tersoff's method.[37] The density isovalue is chosen to be 1e$^{-5}$Bohr$^{-3}$ to give a consistent result with the experiment. The polymers and Au substrate were periodic in the x direction with a lattice constant of 8.64 Å, constituting two anthracene units, or one monomer, for each polymer. A k-point mesh of 5×1×1 was used for the unit cell. The step was along x direction, the distance between periodic images of the step was >40 Å. Four layers of Au atoms were used to model the substrate, with bottom layer atoms fixed. At least 20 Å vacuum spacing was used along the direction perpendicular to the substrate to decouple the periodic image effect. The adsorption energy ($E_{ads}$), calculated as difference of total energy and the sum of energies of two adjacent polymers in vacuum and the Au substrate, is tested for convergence against the energy cutoffs, k-point mesh and vacuum size. Note that the {100}-type step[38] normal to (110) is used in our simulations, which is consistent with our images showing that the {100}-type step is dominant in forming polymers (for example, Fig. 2).

**The relation between dihedral angle and π−π stacking:**

For single polymers, the dihedral angle between neighboring anthracene units remains nearly the same when they change from the adsorption on the terrace (Fig. S2a) to the adsorption on the step edge (Fig. S2b). For two assembled polymers both on the terrace (Fig. S2c), the dihedral angles for the two polymer chains are similar to those in the single polymers. The negligible change of dihedral angles is consistent with the large π−π stacking distance and the small π−π stacking interaction for the CF2-1 configuration. When they are assembled beside the step edge giving CF2-2 configuration (Fig. S2d), the dihedral angles are slightly reduced, which reflects the mediation effect of the step edge. However, the difference of the two angles is small. The dihedral angles show larger differences within the two polymer chains for CF2-3 and CF2-4 configurations with one polymer on the step while the other on the terrace (Fig. S2e,f). The largest difference of dihedral angles is observed for the CF2-4 configuration (Fig. S2f), which shows the shortest π−π stacking distance and the strongest π−π stacking interaction.

**The relation between dihedral angle and cyclodehydrogenation temperature:**

No direct relation is observed between dihedral angle and cyclodehydrogenation temperature. Taking single polymer chains as an example, single polymers respectively adsorbed on the terrace (Fig. S2a) and on the step edge (Fig. S2b) show similar dihedral angles. However, the cyclodehydrogenation temperatures are different, as indicated by the experimental results that the single polymers on the step edge showed an increased threshold of cyclodehydrogenation temperature (see Fig. 3b that shows intact single polymers on the step edge after 570 K annealing). This may be attributed to the strong polymer-substrate interaction when the polymers are adsorbed on the step. On the other hand, for assembled polymer arrays, one may expect a higher cyclodehydrogenation temperature for a larger dihedral angle and a lower temperature for a smaller dihedral angle, such as the two polymers adsorbed on the step and on the lower terrace in the CF2-4 configuration (Fig. S2f). Although the polymer on the lower terrace in CF2-4 shows a comparable dihedral angle to that in single polymers (Fig. S2a), an increased cyclodehydrogenation temperature is experimentally observed for both types of polymers in arrays next to step edges. This can only be explained by the enhanced π−π interaction and thus steric hindrance in the compact assembly configuration.

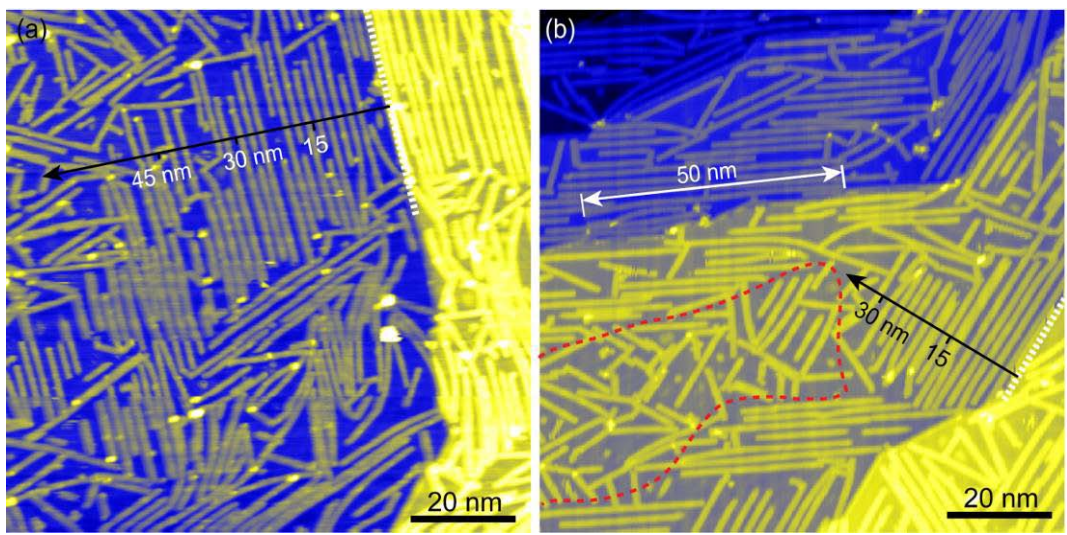

**Fig. S1** (a) Large-area STM image showing well-aligned long GNRs near the straight Au step, while the GNRs become shorter and disordered at regions more than 45 nm away from the step (−2 V, 30 pA). (b) Large-area STM image showing well-aligned long GNRs on narrow terraces and regions close to Au steps, while those away from the steps are short and random (red dashed curves enclosed region) (−2 V, 30 pA). A long GNR with length of about 50 nm is marked near the Au step. A clear decrease of GNR length away from the Au step can be resolved along the black arrow. (a) and (b) are obtained with a precursor deposition pressure of 6×10$^{-9}$ torr, the same as that for Figs. 1 and 2 in the main text.

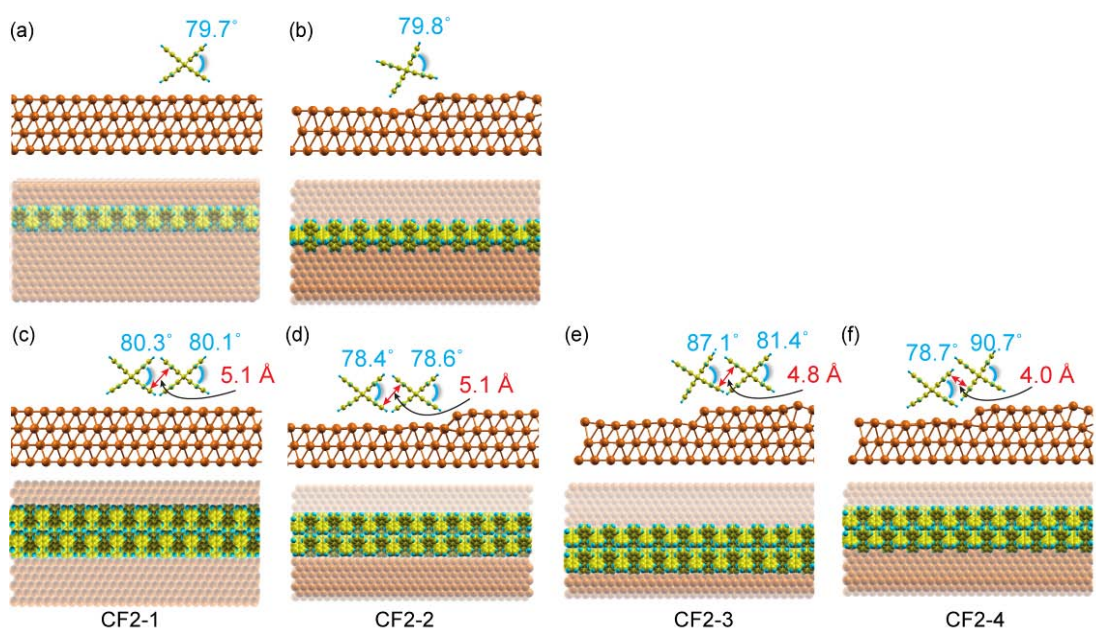

**Fig. S2** (a-b) Dihedral angles between anthracene units in the single polymer chains adsorbed on the terrace and on the step edge, respectively. (c-f) Dihedral angles between anthracene units in the configurations of two assembled polymer chains, the same as those in Fig. 3h in the main text. The angles between anthracene units related to the formation of C-C bonds in the polymers are marked in each panel. The π–π stacking distances in (c-f) are also shown. The top and bottom parts in each panel are the side and top views, respectively.

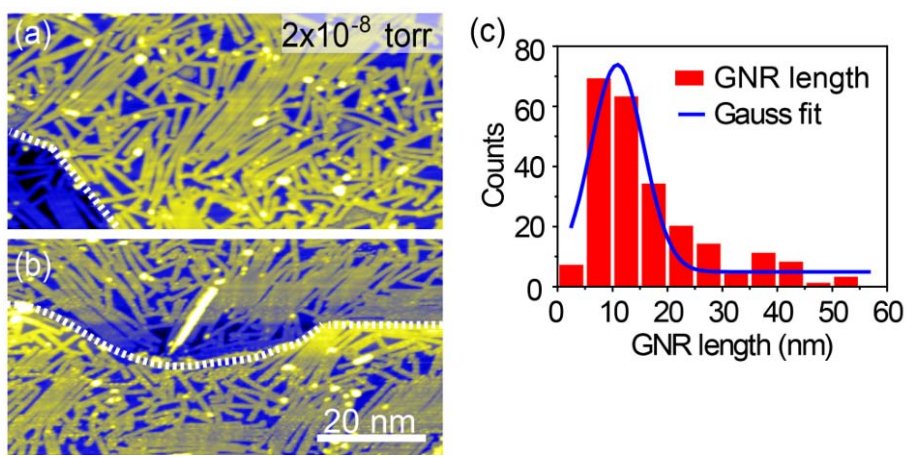

**Fig. S3** (a) and (b) Two representative large-area STM images (both 80×40 nm$^2$) of 7-aGNRs, grown with a high precursor deposition pressure at 2×10$^{-8}$ torr (−3 V, 60 pA). Dashed white lines highlight the Au steps. (c) Length distribution and Gauss fit for the GNRs in (a) and (b). A mean GNR length of about 12 nm is obtained, which is much smaller than that for GNRs near the Au step (~27 nm), grown with a low precursor deposition pressure at 6×10$^{-9}$ torr (Figs. 1, 2 and S1). A low precursor deposition pressure corresponds to a low deposition rate, which gives enough time for molecular precursors to diffuse to and polymerize at the Au step edges, without being blocked by other polymers on the terrace, to achieve well aligned long polymers.

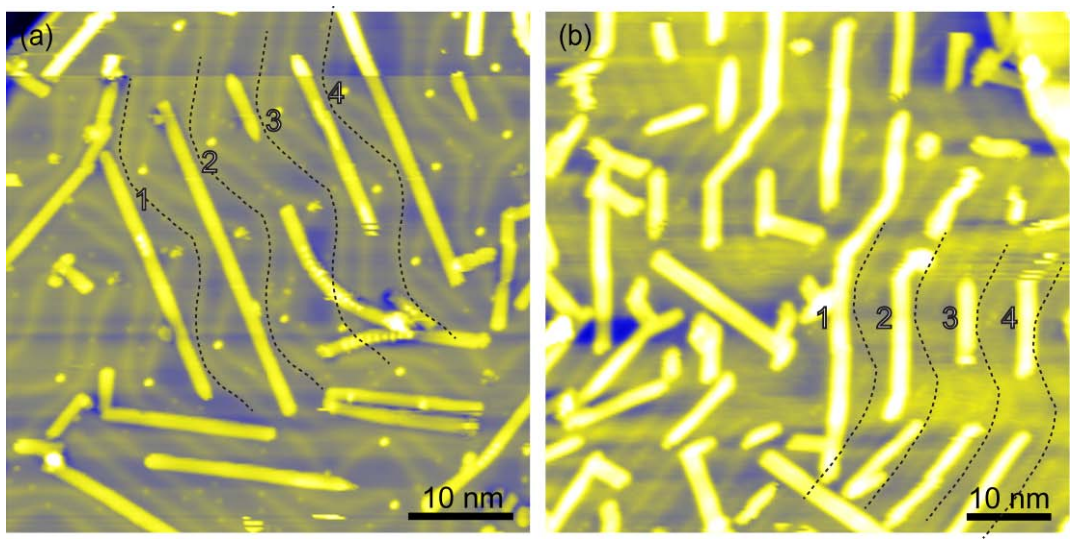

**Fig. S4** (a) and (b) Two representative large-area STM images of low-coverage 7-aGNRs. Setpoint in (a): −2 V, 100 pA; Setpoint in (b): −2 V, 20 pA. Both in (a) and (b), labels 1-4 are provided next to four GNRs growing in parallel and nearly equally separated along the zigzag-patterned (22× √3) herringbone reconstruction, which are similar to previously reported observations.[15]